\documentclass[aps,reprint,superscriptaddress,prl]{revtex4-1}
\usepackage[utf8]{inputenc}
\usepackage[T1]{fontenc}
\usepackage{amsmath,amssymb,graphicx,color}
\graphicspath{{figures/}}
\usepackage[bookmarks=false]{hyperref}
\usepackage[usenames, dvipsnames]{xcolor}
\usepackage{subcaption}
\usepackage[super]{nth}
\usepackage[normalem]{ulem}
\def\be{\begin{equation}}
\def\ee{\end{equation}}
\def\ba{\begin{eqnarray}}
\def\ea{\end{eqnarray}}
\def\CVS{CsV$_3$Sb$_5$~}
\def\KVS{KV$_3$Sb$_5$~}
\def\AVS{AV$_3$Sb$_5$~}
\def\kagome{\text{kagom\'e}~}

\begin{document}

\title{ High Resolution Polar Kerr Effect Studies of CsV$_3$Sb$_5$: Tests for Time Reversal Symmetry Breaking Below the Charge Order Transition}

\author{David R. Saykin}
\thanks{These two authors contributed equally} 
\affiliation{Geballe Laboratory for Advanced Materials, Stanford University, Stanford, CA 94305, USA.}
\affiliation{Department of Physics, Stanford University, Stanford, CA 94305, USA.}
\affiliation{Stanford Institute for Materials and Energy Sciences, SLAC National Accelerator Laboratory, 2575 Sand Hill Road, Menlo Park, CA 94025, USA.}

\author{Camron Farhang}
\thanks{These two authors contributed equally} 
\affiliation{Department of Physics and Astronomy, University of California, Irvine, CA 92697, USA.}

\author{Erik D. Kountz}
\affiliation{Geballe Laboratory for Advanced Materials, Stanford University, Stanford, CA 94305, USA.}
\affiliation{Department of Physics, Stanford University, Stanford, CA 94305, USA.}
\affiliation{Stanford Institute for Materials and Energy Sciences, SLAC National Accelerator Laboratory, 2575 Sand Hill Road, Menlo Park, CA 94025, USA.}

\author{Dong Chen}
\affiliation{Max Planck Institute for Chemical Physics of Solids, 01187 Dresden, Germany.}
\affiliation{College of Physics, Qingdao University, Qingdao 266071, China}

\author{Brenden R. Ortiz}
\affiliation{Materials Department, University of California, Santa Barbara, Santa Barbara, CA 93106, USA.}

\author{Chandra Shekhar}
\affiliation{Max Planck Institute for Chemical Physics of Solids, 01187 Dresden, Germany.}

\author{Claudia Felser}
\affiliation{Max Planck Institute for Chemical Physics of Solids, 01187 Dresden, Germany.}

\author{Stephen D. Wilson}
\affiliation{Materials Department, University of California, Santa Barbara, Santa Barbara, CA 93106, USA.}

\author{Ronny Thomale}
\affiliation{Institut f\"ur Theoretische Physik und Astrophysik, Universit\"at Würzburg, D-97074 Würzburg, Germany.}

\author{Jing Xia}
\affiliation{Department of Physics and Astronomy, University of California, Irvine, CA 92697, USA.}

\author{Aharon Kapitulnik}
\email{aharonk@stanford.edu}
\affiliation{Geballe Laboratory for Advanced Materials, Stanford University, Stanford, CA 94305, USA.}
\affiliation{Department of Applied Physics, Stanford University, Stanford, CA 94305, USA.}
\affiliation{Stanford Institute for Materials and Energy Sciences, SLAC National Accelerator Laboratory, 2575 Sand Hill Road, Menlo Park, CA 94025, USA.}
\affiliation{Department of Physics, Stanford University, Stanford, CA 94305, USA.}

\date{\today}

\begin{abstract}
We report high resolution polar Kerr effect measurements on \CVS single crystals in search for signatures of spontaneous time reversal symmetry breaking below the charge order transition at $T^*\approx 94$ K.  Utilizing two different versions of zero-area loop Sagnac interferometers operating at 1550 nm wavelength, each with the fundamental attribute that without a time reversal symmetry breaking sample at its path, the interferometer is perfectly reciprocal, we find no observable Kerr effect to within the noise floor limit of the apparatus at 30 nanoradians. Simultaneous  coherent reflection ratio measurements  confirm the sharpness of the charge order transition in the same optical volume as the Kerr measurements. At finite magnetic field we observe a sharp onset of a diamagnetic shift in the Kerr signal at $T^*$, which persists down to the lowest temperature without change in trend. Since 1550 nm is an energy that was shown to capture all features of the optical properties of the material that interact with the charge order transition, we are led to conclude that it is highly unlikely that time reversal symmetry is broken in the charge ordered state in CsV$_3$Sb$_5$.
\end{abstract}
\pacs{NaN}

\maketitle

%\section{Introduction}
Symmetry-breaking is the phenomenon where an infinitesimal perturbation can cause the system to break the underlying symmetry of the Hamiltonian. And it is a cornerstone concept in the understanding and manipulation of quantum materials. The state of a system can also be manipulated without explicit symmetry-breaking by controlling topological aspects of the material in momentum space, thereby regulating the electronic band-structure. When strong electron correlations dominate, the system is often observed to be close to multiple competing ordered phases with similar energies. For example, the appearance of charge order may trigger competition with an emerging new electronic state, which in turn leads to the phenomenon of  ``intertwined order'' \cite{Fradkin2015}.

A particularly interesting material system in that respect are the quasi-two-dimensional \kagome compounds \AVS (A=K, Rb and Cs) \cite{Ortiz2019}. These are layered materials with ideal \kagome lattice of V ions coordinated by Sb crystallizing in the P6/mmm space group. Cooling these materials below a temperature $T^*$, the \kagome lattice distorts, undergoing a charge-order transition ($T^*\approx$ 80 K, 110 K and 94 K for K, Rb and Cs respectively) \cite{Ortiz2020,Ortiz2021,Yin2021}, which was shown to be associated with a CDW superstructure modulation displaying chiral anisotropy \cite{Jiang2021,Shumiya2021,Wang2021}. 
Further investigation of the CDW transitions using hard-X-ray scattering revealed that the observed superstructure has in fact a three-dimensional (3D) nature with either $2\times2\times4$ or $2\times2\times2$ superstructure depending on conditions of crystal growth, imposed strain and sample's thermal history \cite{Stahl2022,Ortiz2021Fermi,Liang2021,Li2021,Stahl2022,Xiao2022}.
Intertwined order was initially identified with the discovery of superconductivity in this material system with $T_c\approx2.8$ K for \CVS \cite{Ortiz2020}, where the charge-order transition was first suspected to be wholly electronic in origin.  However, detailed nuclear magnetic resonance (NMR) and nuclear quadrupole resonance (NQR) measurements on CsV$_3$Sb$_5$ single crystals revealed an orbital ordering at $T^*\approx 94$ K clearly induced by a first order structural transition. This is accompanied by electronic charge order that appears to grow gradually below $T^*$ as a secondary (intertwined) order \cite{Song2022}. In fact, a first order characteristic of the transition can be further deduced from the anomalous peak of the specific heat \cite{Ortiz2020,Yu2021,Li2021} and the abrupt large diamagnetic shift of the magnetic susceptibility for magnetic field aligned with the $c$-axis \cite{Ortiz2020,Yu2021}.  The absence of such diamagnetic shift at $T^*$ for magnetic field in the $ab$ plane for \KVS (see ``Extended Data,'' Fig. 3 in \cite{Mielke2022}) and the much reduced effect for \CVS \cite{Chen2022} further support the NMR/NQR findings.

Focusing on the transitions at $T^*$, a key issue of a possible co-occurrence of time-reversal symmetry breaking (TRSB) has emerged through claims of observation of anomalous Hall effect (AHE)\cite{Yang2020,Yu2021,Zhou2022}, changes in muon spin relaxation rate below $T^*$  \cite{YuLi2021,Mielke2022,Khasanov2022}, and magneto-optic Kerr effects  \cite{Wu2021,Xu2022,Hu2022}. From a theoretical standpoint, a CDW formation in kagome metals can appear through both electronically and phonon-mediated mechanisms, or a cooperative version thereof. For \CVS, both electronic and phonon fluctuations seems to be relevant due to the joint presence of van Hove singularities nearby the Fermi level and prominent breathing-phonon modes \cite{Feng2021,Denner2021,Tan2021,Park2021}. The possible appearance of orbital currents has suggested the possibility of a TRSB state below $T^*$, along with an enhanced propensity to nematicity due to multiple CDW nesting vectors \cite{Denner2021,Wulferding2022}. 
Both TRSB and point group symmetry breaking via nematicity are secondary to the translation symmetry breaking and dependent on the precise microscopic setting involving temperature, disorder, and interaction profile.

In this paper we aim to closely examine possible TRSB in high quality crystals of CsV$_3$Sb$_5$ via high resolution measurements of the polar Kerr effect. To substantiate our findings, we use crystals grown in two different laboratories (MPI-Dresden and UC Santa Barbara), all showing salient attributes of the charge order transition at $T^*\approx 94$ K. Further we use two different zero-area-loop Sagnac interferometers (ZALSI) \cite{Xia2006} operating at a wavelength of 1550 nm, one at Stanford University and one at UC Irvine to measure Kerr rotation through the CDW transition, both in zero and in finite magnetic fields. While we clearly observe the abrupt onset of optical birefringence and/or dichroism associated with the structural transition, we see no evidence for a spontaneous Kerr effect below $T^*$ within the volume of the same optical beam to a measurement limit of $\pm 30$ nanoradians, neither in zero-field cool, nor after training with a magnetic field up to 0.34 T. Measurements in an applied magnetic field reveal a diamagnetic Kerr shift, which onsets abruptly below $T^*$ similar to the magnetic susceptibility \cite{Ortiz2020}, but unlike the susceptibility, it stays diamagnetic down to low temperatures. 
\begin{figure}[h]
	\includegraphics[width=.95\columnwidth]{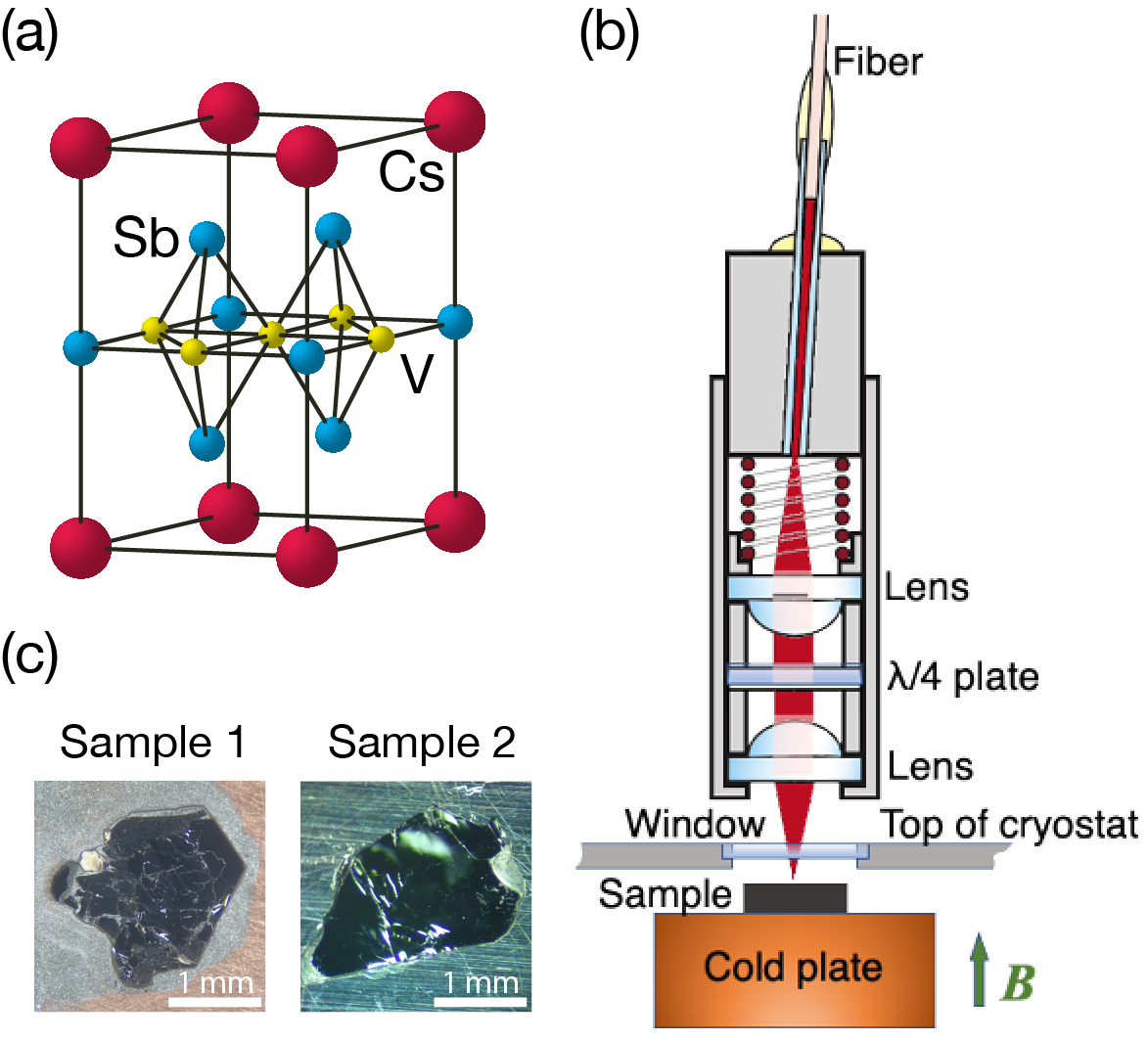}
	\caption{(a) Crystal structure of \CVS. (b) Photo of ``sample 1'' from Dresden, measured at Stanford and ``sample 2'' from UCSB, measured at UC Irvine. (c) Schematic cross-section of probe end optics assembly and sample (not to scale). }
	\label{fig:kerr_data_crystal structure}
\end{figure}

Our results stand in stark contrast to previously published optical measurements claiming finite spontaneous Kerr effects. Two experiments \cite{Wu2021,Xu2022} performed at $800$ nm wavelength measured rotation of linearly polarized light at oblique reflection angle, while the third \cite{Hu2022} used a similar apparatus to our ZALSI at 1550 nm.
We discuss the differences between our measurement and reports \cite{Wu2021,Xu2022} in detail below.
As for the third report \cite{Hu2022}, we speculate that a series of uncontrolled subtractions of data may have been the source of a false positive result.

\begin{figure*}[!ht]
	\centering
	\includegraphics[width=\textwidth]{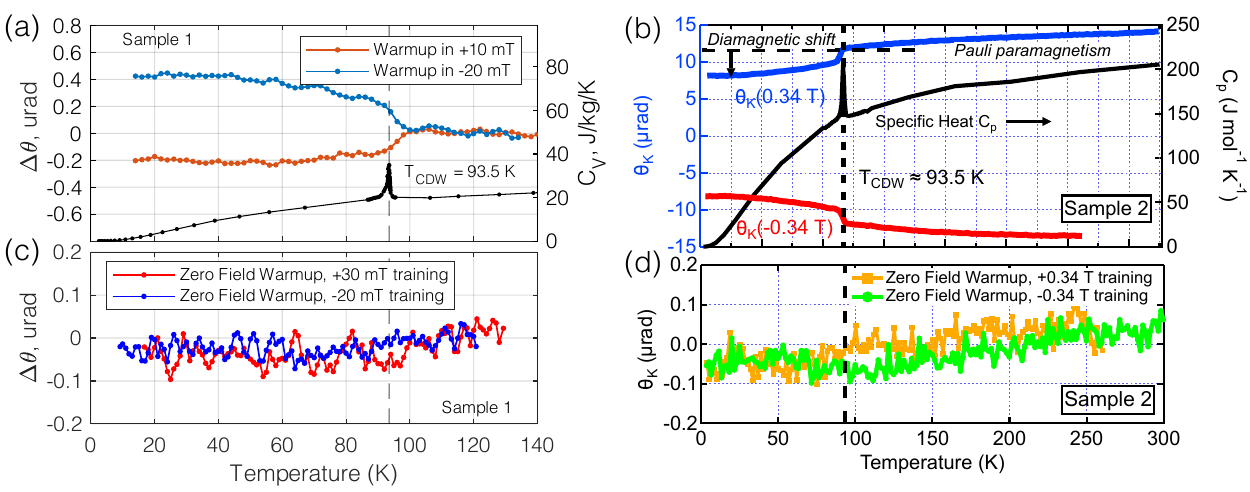}
	\caption{\textbf{Kerr signal in CsV$_3$Sb$_5$}. 
	(a) Sample 1 measured at Stanford: Relative Kerr angle $\Delta \theta = \theta - \theta(120 K)$ (left axes) in external magnetic fields of 20 mT directed along $\pm\hat{z}$, showing the onset of diamagnetic shift when the specific heat $C_V$ (right axes) develops a peak at $T^*\approx 94$ K. (b) Sample 2 measured at Irvine: Absolute Kerr signal $\theta_K$ (left axes) during $\pm$0.34 T cooldowns showing the same onset of diamagnetic shift that coincides with the anomalous peak in specific heat $C_p$ (right axes). (c) Sample 1: Kerr angle $\Delta \theta$ measured during zero field warmups after +30 mT and -20 mT trainings. (d) Sample 2:  Kerr signal $\theta_K$ measured during zero field warmups after $\pm$0.34 T trainings. There is no discernible spontaneous Kerr effect below $T^*$ to within $30$ nrad in either samples. }
	\label{fig:kerr_data}
\end{figure*}

%\section{Results}
%\subsection{Samples}
In this study we use high quality and well-characterized single crystals of \CVS obtained from two different sources. Samples grown at UCSB (dubbed sample 2) followed the previously published procedure found in \cite{Ortiz2020} and related publications, while samples grown at MPI-Dresden (dubbed sample 1) followed the previously published procedure found in \cite{Huang2022} and related publications. The crystal structure of \CVS and photos of two example crystals from the two sources are presented in Fig.~\ref{fig:kerr_data_crystal structure}. Specific heat data obtained on these crystals show the ``standard'' sharp first--order--like transition peak at 93.5 K  as is shown in Fig.~\ref{fig:kerr_data}. This charge density transition is also marked by the kink in DC resistance and the pronounced peak in its first derivative $dR/dT$ as shown in Supplemetary Figure \cite{supp}.
Moreover, we observe two additional features in the $dR/dT$ curve. There is a dispersive line shape close to the temperatures where a kink was observed in NMR $1/T_1T$ \cite{Song2022} and where the electronic magneto-chiral anisotropy (eMChA) was observed in nonlinear transport \cite{Guo2022}, suggesting an additional phase transition.
Also there is a small bump in resistance derivative happening at temperatures below $T^*$ which could be related to either onset of CDW along $c$ direction or to formation of $4a_0$ stripe order \cite{Li2023}.
Importantly, we also detect onset of charge order simultaneously with Kerr effect measurements within the same optical volume through reflectivity and coherent reflection ratio measurements.

%\subsection{Optical Measurements}
High resolution Polar Kerr effect measurements presented in this paper were performed using zero-area-loop Sagnac interferometers (ZALSI) first introduced by Xia {\it et al.} \cite{Xia2006}. By construction the ZALSI is inherently reciprocal and thus by its symmetry it can distinguish between true TRSB Kerr effect and optical activity when reflected from a chiral medium.
The light beam of ZALSI is focused onto an optically flat region that can be hundreds of microns in size. One such region is illustrated in the scanning images of reflected power $P_0$ on sample 2 at $3.1 K$ provided in supplementary materials \cite{supp}. In a magnetic field of -0.34 T the Kerr signal is uniform across this flat region.  The temperature dependence of the in-field Kerr signals at fixed spots are shown in Fig.~\ref{fig:kerr_data}A for sample 1 measured at Stanford, and in Fig.~\ref{fig:kerr_data}B for sample 2 measured at Irvine, alongside the measured specific heat. In both cases, there is a clear and relatively abrupt onset of diamagnetic shift that coincides with the anomalous peak in the specific heat at $T_\text{CDW}$, confirming the CDW transition in the probed optical volume.

The main result of this work is the test for spontaneous Kerr effect in \CVS. Kerr signal was recorded during zero field warmup after removal of a ``training'' magnetic field applied during cooldown. Results from the two laboratories are presented in Fig.~\ref{fig:kerr_data}C and Fig.~\ref{fig:kerr_data}D respectively, showing no discernible Kerr effect below $T^*$ to within 50 nrad. This uncertainty is a factor of $\sim1000$ smaller than the recent MOKE measurements \cite{Wu2021, Xu2022, Hu2022}. Similarly, no spontaneous Kerr signal was found during zero field cooldowns and subsequent zero field warmups. The absence of spontaneous Kerr signal is also universally found across the sample surface, which can be seen in the scanning Kerr signal image presented in the supplementary materials \cite{supp} taken on sample 2 at $3.1 K$ after removal of the -0.34 T magnetic field.

%\section{Discussion}
Our main result raises doubts to whether time reversal symmetry is indeed broken below $T^*$ in  \CVS. On the other hand, measurements at a finite magnetic field oriented along the $c$-axis show a clear and relatively abrupt diamagnetic shift in the Kerr effect, which near $T^*$ behaves similarly to the in-field magnetic susceptibility measurements \cite{Ortiz2020,Yu2021,Wang2021}. However, the detected average (DC) optical power ($P_0$) and the coherent reflection ratio ($P_2/P_0$) (see supplementary material \cite{supp}) measurements in either zero field or in-field, which are taken simultaneously with the Kerr measurements within the same optical volume, consistently show an abrupt transition at $T^*$ with a width of $\sim2$ to 3 K, irrespective of the applied fields 
(similar result is reported in \cite{Zhou2021,Uykur2021}).
Since TRSB effect is expected to be secondary to any structural or CDW transitions, and grow gradually below $T^*$ \cite{Denner2021}, it is unlikely that the observed abrupt changes in the differential reflectivity are associated with such a TRSB effect.  A similar scenario was previously observed in the Cuprate superconductors where a first-order structural transition from an orthorhombic to tetragonal phase at $T_s$ (so called LTO to LTT transition) in La$_{2-x}$Ba$_x$CuO$_4$ \cite{Hucker2011} was accompanied by a charge-order transition with an order parameter that evolved gradually below $T_s$. 

The bulk magnetic susceptibility of \CVS has been previously shown to develop a paramagnetic response below $\sim50$ K \cite{Ortiz2020,Yu2021,Wang2021}, presumably due to bulk impurities, overwhelming the initial diamagnetic drop below $T^*$.
The Sagnac interferometers probe only the freshly cleaved surface region within a small volume defined by the beam diameter $d=10$ $\mu$m and the optical penetration depth $\delta\sim 20$ nm, and thus are less susceptible to these bulk impurities, particularly as we typically use a lower magnetic field. Hence, the change in the Kerr signal below $T^*$ stays diamagnetic and is almost temperature independent as seen in the relative Kerr angle $\Delta \theta=\theta-\theta(120 K)$ for sample 1 in Fig.~\ref{fig:kerr_data}A and in the absolute Kerr signal $\theta_K$ for sample 2 in Fig.~\ref{fig:kerr_data}B. 

The susceptibility measurements on \CVS further show that the diamagnetic drop with the field along the $ab$ plane is almost 5  times weaker than if the field points in the $c$-direction, where earlier results on \KVS show no diamagnetic shift for the field along the $ab$-plane \cite{Ortiz2020}. These observations indicate that the diamagnetic response originates from in-plane effect, presumably orbital effects such as non-interacting loop currents as suggested by the NMR/NQR experiments \cite{Song2022}. In contrast, the paramagnetic response observed in magnetic susceptibility is isotropic with respect to the applied magnetic field and remains on the scale of the Pauli paramagnetism above $T^*$. Hence, we conclude that we have no evidence for spontaneous ferromagnetic orderings of loop currents.
However, our results still allow for an antiferromagnetic ordering of loop currents.

Next we provide our speculations on why results reported here are in contrast to other optical measurements that attempt to measure Kerr effect in the \AVS systems \cite{Wu2021,Xu2022}.
First of all, 
we note that while the ZALSI is designed to only detect TRSB effects in the sample as an excess signal beyond the natural reciprocity of the apparatus, other techniques may need to employ less controlled means to scrutinize true TRSB effects. 
Secondly, 
we note that measurements \cite{Wu2021,Xu2022} have been made at oblique incident angle, while we have measured light reflected at normal incidence. It is known that,
in an optically active material, the polarization of reflected light at oblique incidence will always rotate with respect to the initial state of polarization irrespective of orientation between initial polarization and principal axes of the crystal \cite{Silverman1987,Silverman1992}. Furthermore, the statement that no finite optical rotation is possible for reflected light at normal incidence is true for polar Kerr effect, but does not apply to circular dichroism, which originates from different optical penetration depths for the two circularly polarized components.
Thirdly,
if circular dichroism is detected at normal incidence, it is not a proof for TRSB. In fact, in most cases it will indicate circular birefringence that respects time reversal symmetry. A notable example is reflection from the surface of tellurium, where large rotary power and circular dichroism were previously measured \cite{Fukuda1975,Ades1975,Stolze1977}, but no Kerr effect was detected using our ZALSI \cite{Fried_Thesis}. 
Lastly, 
concerning the difference in photon wavelength used in our system and in \cite{Wu2021,Xu2022}, we turn to the optical properties of \CVS \cite{Zhou2021, Uykur2021, Uykur2022} and note that our wavelength is close to the Lorenz peak in optical conductivity centered around $0.6$ $\mu$m${}^{-1}$ wavenumbers (dubbed L${}_1$ in \cite{Zhou2021}), which is enhanced by CDW. Hence, we would expect that our wavelength probes exactly the electronic states responsible for charge order formation, unlike the $\sim800$ nm wavelength which is further away from the CDW energy scale.
Thus, we conclude that unless there is a special reason for Kerr rotation to zero at 1550 nm, which is unlikely in a good metal, our wavelength should provide a more sensitive probe compared to the one used in \cite{Wu2021,Xu2022}.

The above discussion suggests that at least within the technical aspect, those other optical measurements fall short in comparison to the ZALSI, which inherently measures only non-reciprocal effects, thus returning a zero signal if time reversal symmetry is not broken.
On the materials side we note that if the CDW is chiral along the $c$-axis of the material, mirror symmetries are broken as was previously observed in STM \cite{Jiang2021,Shumiya2021,Wang2021}, and circular dichroism is allowed. Moreover, a finite minimum value of polarization rotation will be present with magnitude that depends on the pitch of the chiral CDW, which depends on the initial symmetry of the unbroken state, as well as the optical penetration depth at the probing photon energy.  These effects may then be incorrectly interpreted as signatures of TRSB. 
From the technical point of view we also note that using linear polarization to probe complex materials requires perfect alignments of the optical components with respect to the plane of incidence. For example, if rotation of the incident light polarization is performed using a perfect half-wave plate (HWP), which is tilted by a small angle with respect to the propagation axis, even a perfect metallic mirror will show an apparent polarization rotation \cite{Zhu1994}. For \CVS parameters, a 5 degrees error in the HWP alignment yields a $\pm1$ millirad error. Commercial zero-order HWPs have a typical retardance error of 1$\%$ even for normal incidence, yielding further errors. 

Finally we briefly comment on the reported onset of anomalous Hall effect \cite{Yang2020,Yu2021,Zhou2022} and change in muon spin relaxation below $T^*$  \cite{YuLi2021,Mielke2022,Khasanov2022}. Indeed, the anomalous Hall effect is a direct measure of ferromagnetic order. However, in general it implies a finite value at zero applied magnetic field. Yet, all reports of AHE in \CVS show no Hall resistance at zero magnetic field. In that respect this AHE is not strictly related to spontaneous TRSB effect and may come from non-TRSB chiral effects \cite{Konig2019} or change in carrier density and mobility due to charge order.  
Another set of experiments which are at odds with our results involve $\mu$-Sr measurements, and in particular the study in \cite{Mielke2022} where a TRSB is inferred from a rather small change through a sharp onset in muons relaxation below $T^*$ at zero field, followed by an almost temperature independent trend at low temperatures. Indeed, as a purely magnetic probe, $\mu$Sr is not directly sensitive to charge order and thus a change in relaxation might be interpreted as evidence for TRSB in the electronic system. However, $\mu$Sr does sense the presence of charge order via the magnetic dipolar coupling of the muons with the host nuclei at zero or low magnetic fields (see e.g. \cite{Sonier2014}).  In particular, since the zero and low field effect in the $\mu$Sr studies is abrupt and follows in trend the structural transition \cite{Song2022}, it is likely to induce a change in the muons' preferred ``rest positions'' above and below $T^*$, which will show up as change in relaxation rates. While an all-electronic spontaneous TRSB buildup mechanism for the CDW \cite{Denner2021} is in principle not inconsistent with the $\mu$Sr data deeper in the CDW phase, the abruptness in the $\mu$Sr response at the transition cannot solely be reconciled from this angle, and suggests a pivotal role of phonon-mediated CDW formation related to a structural first order transition. Furthermore, the in-field enhancement of the signal in magnetic field seems to follow the NMR results, which again cannot directly point to a spontaneous TRSB.

%\subsection{Conclusions}
In conclusion, we have used high resolution polar Kerr effect measurements to scrutinize previously reported submissions \cite{Wu2021,Xu2022,Hu2022} that time reversal symmetry is spontaneously broken below the charge order transition a in \CVS (marked as $T^*$).  For these studies we used two different variation of a zero-area loop Sagnac interferometer operating at 1550 nm wavelength, each with the fundamental attribute that without a time reversal symmetry breaking sample at its path, the interferometer is perfectly reciprocal and thus yield a zero Kerr signal at the detector to within its noise floor limit of $\sim30$ nanoradians. We then show that high quality \CVS single crystals obtained from two different crystal-growth laboratories show no resolvable Kerr signal at zero magnetic field to within the instruments noise floor level, whether the sample was cooled at zero field, or was trained through $T^*$ at a finite field and then measured at zero field upon warming up. Concurrent measurements of the coherent reflection ratio show a sharp transition within the same optical volume where Kerr effect was measured, thus supporting the robustness of our measurements. At finite magnetic field we observe a sharp diamagnetic transition at $T^*$, similar to the transition observed in magnetic susceptibility, except that the Kerr data remain diamagnetic down to low temperatures without change in trend. Since 1550 nm is an energy that was shown to capture all features of the optical properties of the material that interact with the charge order transition, we are led to conclude that it is highly unlikely that time reversal symmetry is broken in the charge ordered state in \CVS.

\paragraph{Note added:} during the review process of this manuscript, we've learnt that recent measurement \cite{Farhang2023} demonstrated that rotation of linear polarization in \CVS has isotropic component $\theta_C$ even at normal incidence that is comparable in size to those found at 800 nm \cite{Wu2021,Xu2022}. Moreover, it was found that $\theta_C$ does not respond to applied magnetic field and changes sign and magnitude from sample to sample. These suggest that the polarization rotations reported in \cite{Wu2021,Xu2022} may not be related to MOKE.

\section{Acknowledgements}
Work at Stanford University was supported by the U.S. Department of Energy, Office of Science, Basic Energy Sciences, Division of Materials Sciences and Engineering, under Contract DE-AC02-76SF00515. Work at UC Irvine was supported by the Gordon and Betty Moore Foundation through Grant GBMF10276. S.D.W. and B.R.O. acknowledge support via the UC Santa Barbara NSF Quantum Foundry funded via the Q-AMASE-i program under award DMR-1906325. R.T. acknowledges support from the Deutsche Forschungsgemeinschaft (DFG, German Research Foundation) through QUAST FOR 5249-449872909 (Project P3), through Project-ID 258499086-SFB 1170, and from the Würzburg-Dresden Cluster of Excellence on Complexity and Topology in Quantum Matter – ct.qmat Project-ID 390858490-EXC 2147.

\bibliography{kerr}

\appendix*

\section{Supplementary Materials}

\section{Samples}\label{app:samples}

\setcounter{equation}{0}\renewcommand{\theequation}{A.\arabic{equation}}
In the present study two \CVS samples grown in different laboratories were measured in two Sagnac interferometers. First samples dubbed "sample 1" is from Cluadia Felser group in Dresden, it was measured in Kapitulnik group at Stanford, second sample dubbed "sample 2" grown at Stephen D. Wilson's group at UCSB was studied at UC Irvine in Xia group. 

\section{Zero-area-loop Fiber-optic Sagnac Interferometer (ZALSI)}\label{app:ZALSI}
\setcounter{equation}{0}\renewcommand{\theequation}{A.\arabic{equation}}

Samples were measured using zero loop area fiber Sagnac interferometers (ZALSI) \cite{Xia2006} using $30$ $\mu$W optical power at 1550 nm wavelength with phase modulation at $\omega=5$ MHz. Two low-coherence light waves of right and left circularly polarizations were sent to the sample. And the non-reciprocal phase difference $\varphi_{nr} = 2 \theta_K$ between the two lights acquired upon reflection was detected with lock-in amplifiers. By construction, unlike a standard ellipsometer, this approach fundamentally rejects polarization rotations due to non-TRSB effects such as linear and circular birefringence and dichroism that could mimic a TRSB Kerr signal. In addition, by reducing the Sagnac loop to zero area within a single fiber, it also rejects a background Sagnac signal from earth rotation, which breaks time-reversal symmetry and is the basis for fiber gyroscopes. 

\begin{figure}[h]
	\includegraphics[width=\columnwidth]{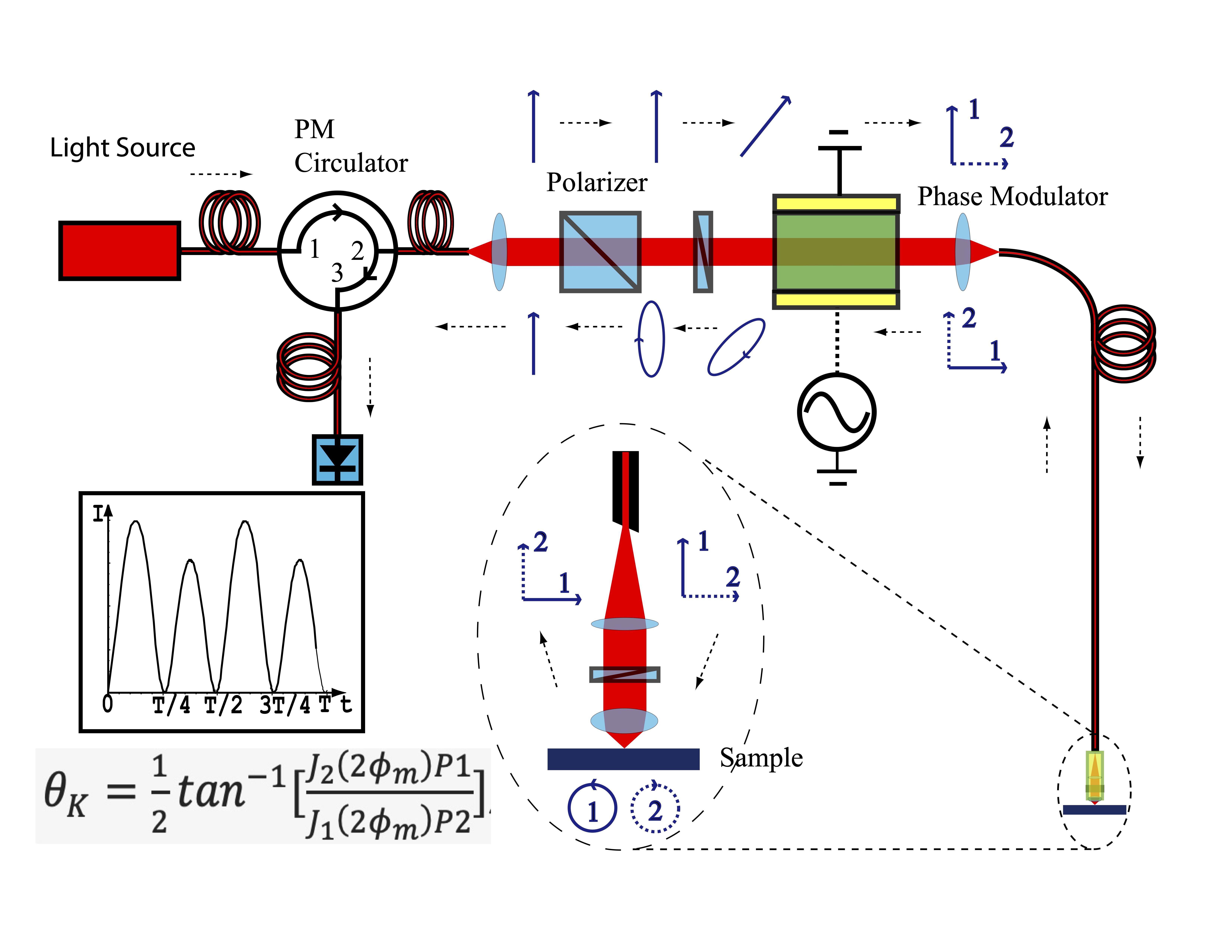}
	\caption{\textbf{ZALSI.} Schematics of the zero-area-loop fiber-optic Sagnac interferometer (ZALSI).}
\label{fig:ZALSI}
\end{figure}

\begin{figure*}[!ht]
	\centering
	\begin{subfigure}{.46\linewidth}
		\label{fig:refl_data_Stanford}
		\includegraphics[width=\textwidth]{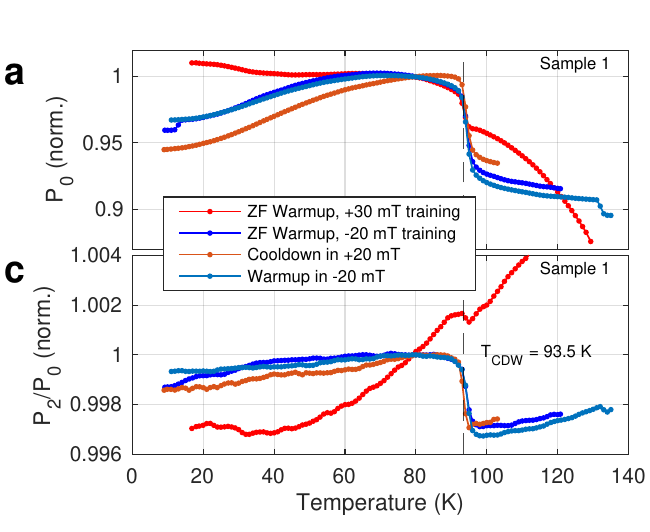}
	\end{subfigure}
	\quad
	\begin{subfigure}{.47\linewidth}
	\vspace*{3mm}
		\label{fig:refl_data_Irvine}
		\includegraphics[width=\textwidth]{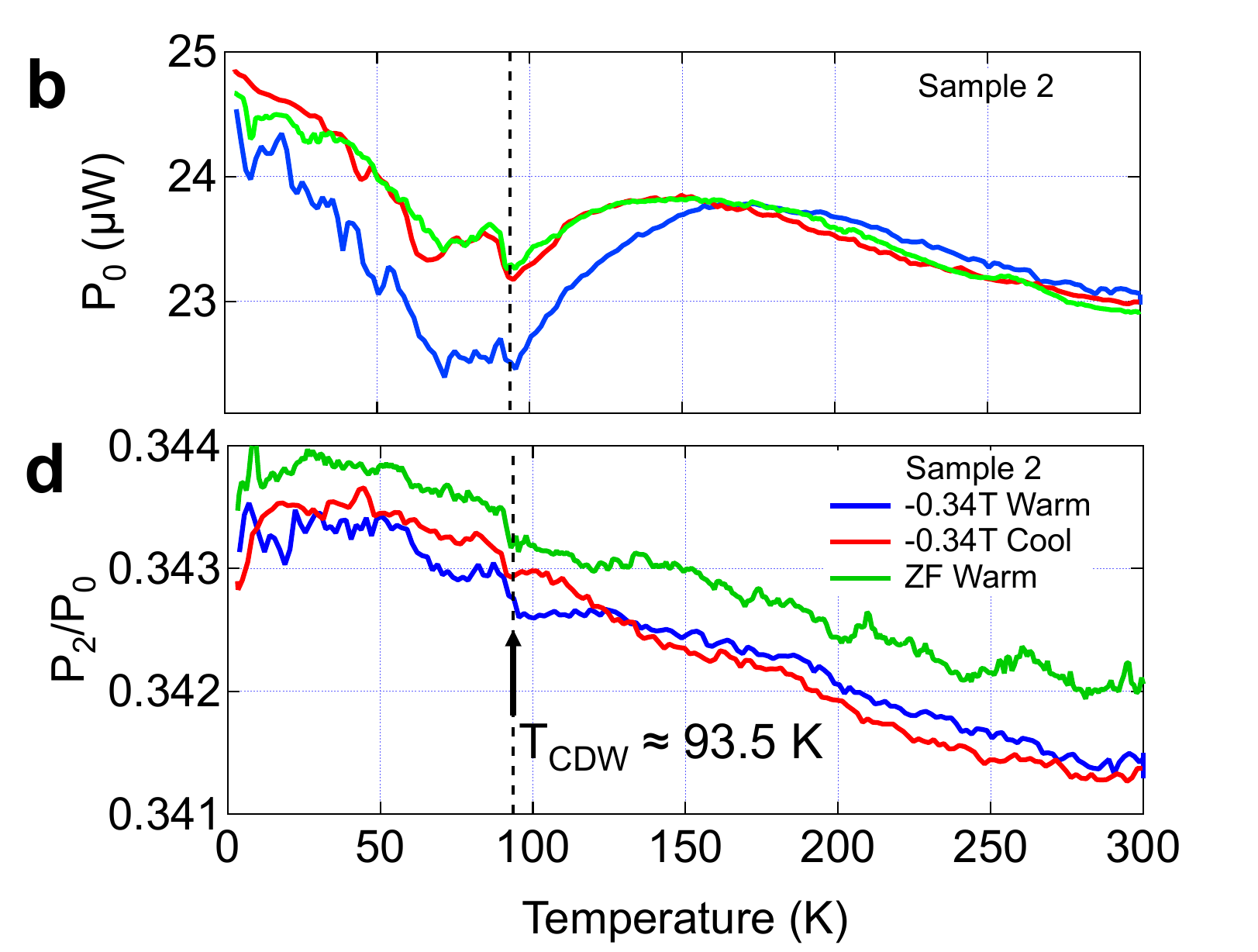}
	\end{subfigure}
	\caption{\textbf{Birefringence and dichroism in \CVS detected by Sagnac interferometers}. \textbf{a}, Normalized average (DC) power $P_0$ in sample 1. \textbf{b}, Average (DC) power $P_0$ in Sample 2. \textbf{c}, Coherent reflection ratio $P_2/P_0$ in sample 1. \textbf{d}, $P_2/P_0$ in sample 2. The abrupt change of $P_2/P_0$ at $T^*\approx 94$ K in both samples indicate the onset of birefringence and/or dichroism (linear or circular) in \CVS.}
	\label{fig:refl_data}
\end{figure*}

The apparatus is based on the Sagnac interferometer \cite{Sagnac1913} where two counter-propagating beams travel the exact same optical path before interfering at the detector. In our implementation we use two orthogonally linearly polarized light beams, which are then converted into right and left circular polarizations using a quarter-wave plate. An objective lens focuses the light onto a small interaction region on the sample, and the reflected light beams are converted back to linear polarizations with exchanged polarization directions. Owing to the reciprocity of the apparatus, a non reciprocal phase shift $\phi_{nr}=2\theta_K$ will appear at the detector only if time reversal symmetry is broken through the interaction of the two circularly polarized beams with the sample \cite{Xia2006} (see supplementary material \cite{supp}). With careful construction we can routinely achieve shot-noise-limited senstivity of ($\sim$100 nrad$/\sqrt{Hz}$) for optical power above $\sim 10 \ \mu$W, with magneto-optic Kerr and Faraday resolutions of nrad limited by the long-term drift of the instrument. Some of our notable accomplishments include the study of TRSB in Sr$_2$RuO$_4$ \cite{PXia2006}, UPt$_3$ \cite{Schemm2014} and more recently UTe$_2$ \cite{Hayes2021}, elucidating the limit for ferromagnetism in thin SrRuO$_3$ films \cite{Xia2009} and discovery of the inverse proximity effect in ferromagnet/superconductor bilayers \cite{Pxia2009} to name a few. Again, to be able to reliably scrutinize the possible spontaneous polar Kerr effect below $T^*$, measurements were performed in two different apparatuses, one at Stanford University and one at UC Irvine. While both systems were constructed in a similar way employing a broadband superluminescent light emitting diode (SLED) with a center wavelength of 1550 nm, the details of the sample holder, windows construction and end-optics are slightly different. 

A schematics of ZALSI as well as polarization states at each point are shown in Fig.~\ref{fig:ZALSI}. The beam of light polarized at 45$\rm ^o$  to the axis of a electro-optic modulator (EOM), which generates $5$ M\rm Hz time-varying differential phase shifts $\phi_m$ along its two major axis and split the light into two beams of roughly equal powers. The two beams are then launched into the fast and slow axes respectively of a polarization maintaining (PM) fiber. Upon exiting the fiber, the two orthogonally polarized beams are converted into right- and left-circularly polarized light by a quarter-wave ($\lambda /4$) plate, and are then focused onto the sample. The non-reciprocal phase shift $\phi_{nr}$ between the two circularly polarized beams upon reflection from the sample is twice the Kerr rotation ($\varphi_{nr} = 2 \theta_K$). The same quarter-wave plate converts the reflected beams back into linear polarization, but with a net 90$^o$ rotation of the polarization axis. The two beams then travel back through the PM fiber and the EOM with exchanged axes before they arrive again at the polarizer. At this point, the two beams have gone through exactly the same path but in opposite directions, except for a phase difference of  $\phi_{nr} = 2 \theta_K$ from reflection off of the sample. The two beams are once again coherent, and interfere to produce an elliptically polarized beam, whose in-plane component is routed by the circulator to the photodetector. Lock-in detection was used to measure the average (DC) power ($P_0$), the first harmonics ($P_1$), and the second harmonics ($P_2$) of the detected optical power $P(t)$:

\begin{equation}
\centering
P(t)=\frac{1}{2}P[1+\cos(\varphi_{nr}+2\phi_m\sin(\omega t))]
\end{equation}

where $P$ is the returned power without modulation, and depends on focus of the objective lens and sample reflectivity. $P(t)$ can be further expanded into Fourier series of $\omega$ if we keep $\varphi_{nr}$ as a slowly time-varying quantity compared to $\omega$:

\begin{align}
\centering
	\begin{aligned}
		P(t)/P &= [1+J_0(2\phi_m)]/2 \nonumber\\
			&+ (\sin (\varphi_{nr}) J_1(2\phi_m))\sin(\omega t) \nonumber\\ 
			&+ (\cos (\varphi_{nr}) J_2(2\phi_m))\cos(2\omega t)\nonumber\\
			&+ 2J_3(2\phi_m)\sin(3\omega t)\nonumber + ... 
	\end{aligned}
 	\label{eq:fourier}
\end{align}

where $J_1$ and $J_2$ are Bessel functions. Therefore, the detected powers $P_0$, $P_1$ and $P_2$ are: 

\begin{align}
	\centering
		P_0/P & =[1+J_0(2\phi_m)]/2,	\\
		P_1/P &= (\sin (\varphi_{nr}) J_1(2\phi_m)), \\
		P_2/P &= (\cos (\varphi_{nr}) J_2(2\phi_m))
\end{align}
Hence Kerr signal $\theta_K=\varphi_{nr}/2 $ can be obtained using equation \ref {eq:Sagnac_kerr}, which is independent of optical power, sample reflectivity and focus of the objective lens. For optimal $\theta_K$ sensitivity $\phi_m$ is often chosen to be close to 0.92. 

\begin{equation}
\theta_K = \frac{1}{2} \tan^{-1}\left[ \frac{J_2(2\phi_m)P_1}{J_1(2\phi_m) P_2}\right]
\label{eq:Sagnac_kerr}
\end{equation}

\section{Time-reversal Symmetry Preserving design}

Here we provide a qualitative argument on why ZALSI only detects a signal when TRS is broken. Assume a sample with its surface in the $x$-$y$ plane and a linearly polarized electric field propagating along the $z$-direction which is perpendicular to the surface of the sample.  Adopting the common convention where the sense of circular polarization is determined with respect to a fixed axis (here the $+z$ axis), the handedness of the reflected light with respect to the direction of propagation flips.  This convention defines four different indices of refraction:  $n_{R\nearrow}$ and $n_{L\nearrow}$ are the refractive indices for right circularly polarized (RCP) light and left circularly polarized (LCP) light propagating along $+z$ (reflected light) and  $n_{R\searrow}$ and $n_{L\searrow}$ are the refractive indices for RCP and LCP propagating along $-z$ (incident light).  
Where TRS is preserved 
\begin{align}\label{eq:TRS_n_RL}
    \begin{aligned}
        n_{R\nearrow} &= n_{L\searrow},	\\
        n_{L\nearrow} &= n_{R\searrow},
    \end{aligned}
\end{align} 
however, if TRS is broken at least one equality in \eqref{eq:TRS_n_RL} fails. 
Focusing on a polar configuration where light impinges on the sample at normal incidence (as it is in our ZALSI apparatus) and using scattering theory to calculate the Kerr effect from the difference between the reflection amplitudes for right and left circularly polarized light \cite{Kapitulnik2015}, it can be shown that 
\begin{align}
    \theta_K\propto[ ( n_{R\nearrow} + n_{R\searrow}) - ( n_{L\nearrow} + n_{L\searrow} ) ].
\end{align}
It is then clear that $\theta_K$ is finite \emph{only} for the case of TRSB. The above results are robust, independent of whether the material that we study is isotropic, or it possesses linear birefringence or natural optical activity. 

\begin{figure}[h]
	\includegraphics[width=\columnwidth]{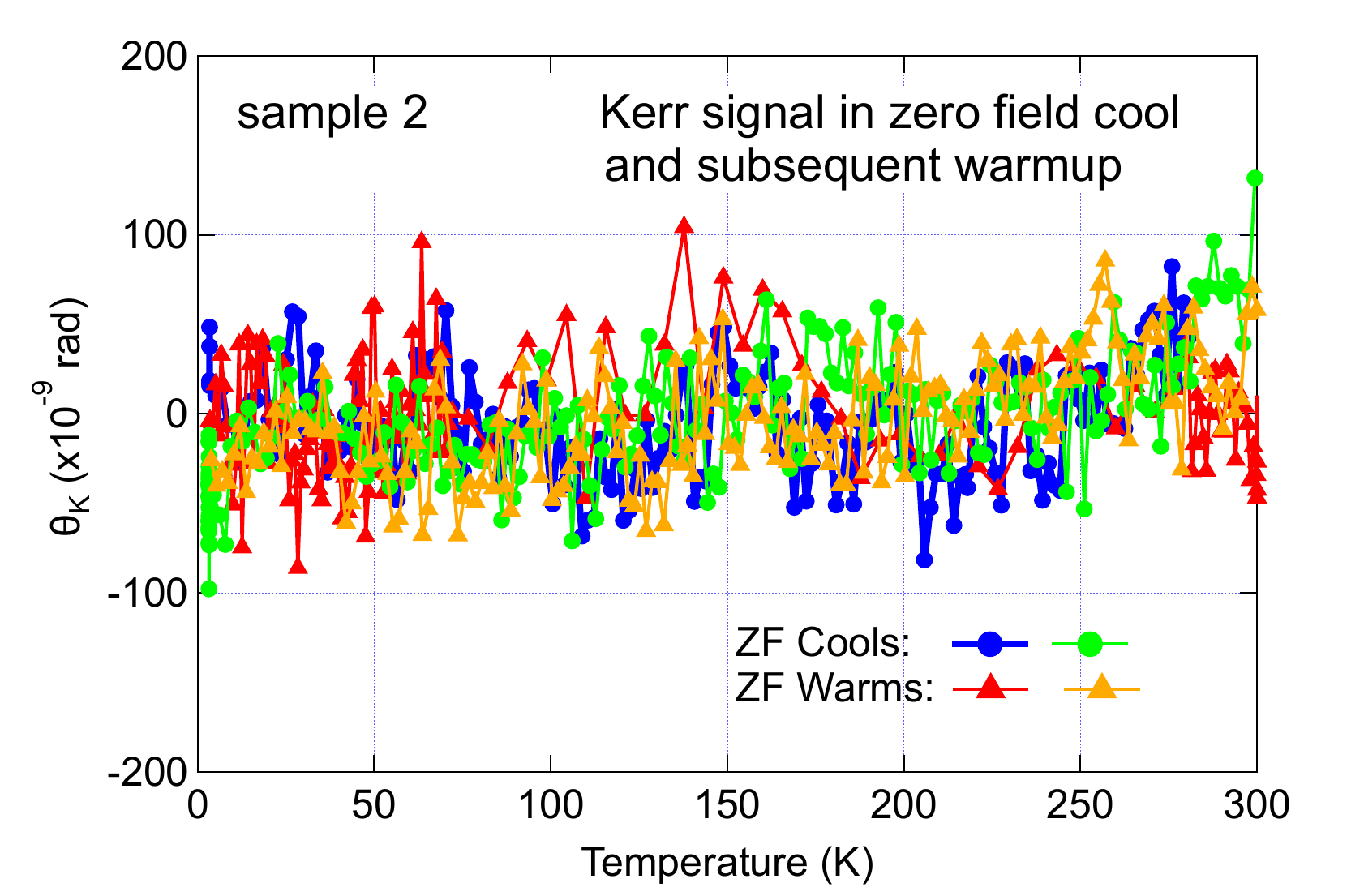}
	\caption{\textbf{Zero-field cool.} Kerr signal during zero-field cools and subsequent warmups in sample 2 measured at Irvine, showing no spontaneous Kerr signal above $30$ $nrad$.}
\label{fig:ZFCool}
\end{figure}

\begin{figure}[h]
	\includegraphics[width=\columnwidth]{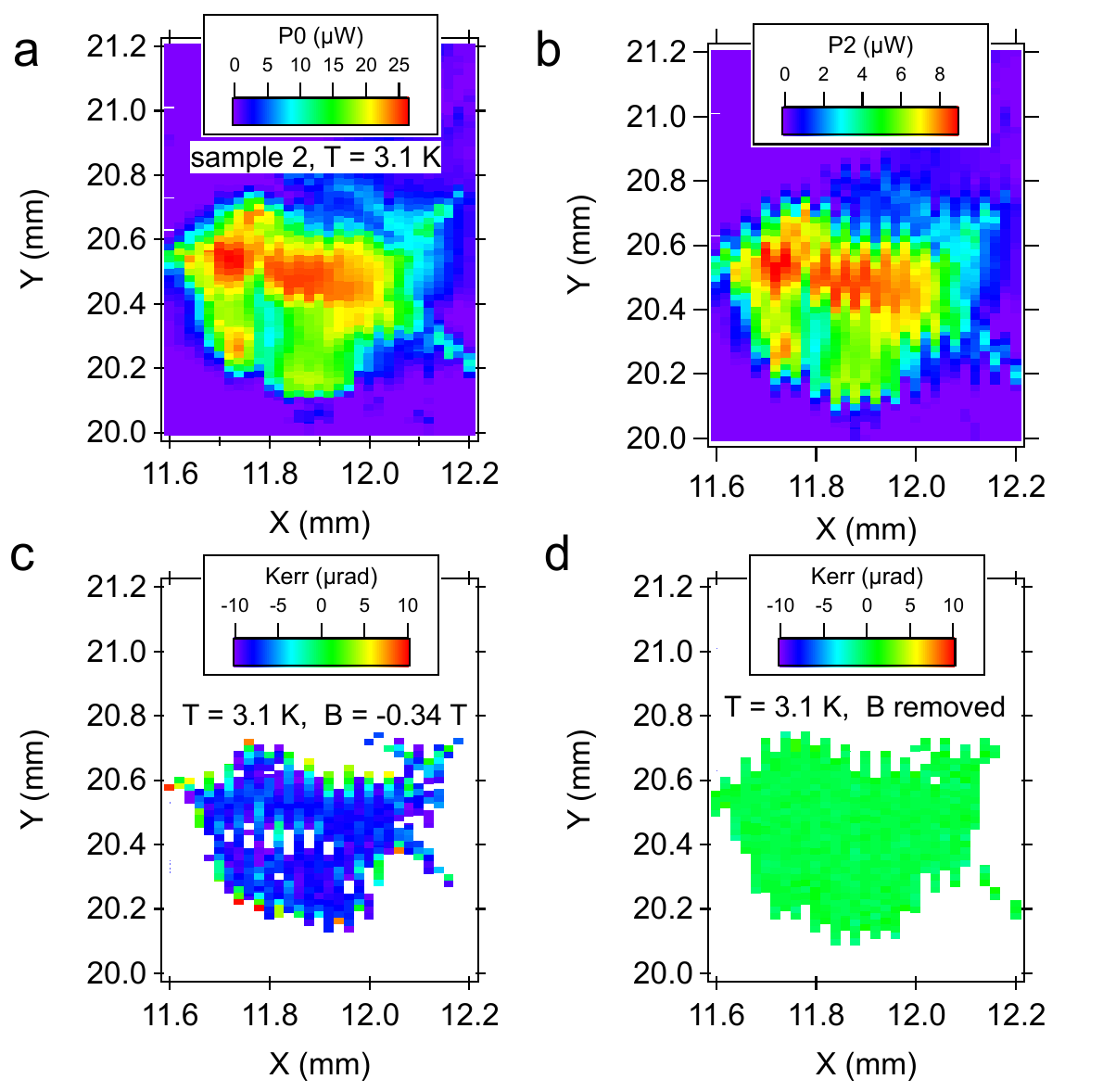}
	\caption{\textbf{Sagnac scanning images.} Sample 2 measured at Irvine. \textbf{a}, P0 and \textbf{b}, P2 are detected DC and second harmonics optical powers taken at T = 3.1 $K$ sample temperature, showing an optically flat region colored in red to green. \textbf{c}, Kerr image taken at T = 3.1 $K$ and in a magnetic field of B = -0.34 T, showing a uniform Kerr signal of - 8.16 $\ \mu$rad (-24 $\ \mu$rad/T). \textbf{d}, Kerr image taken after the magnetic field is removed, showing zero spontaneous Kerr signal.  }
\label{fig:scan}
\end{figure}

\section{Detection of CDW with coherent reflection ratio $P_2/P_0$}\label{app:jones}
\setcounter{equation}{0}\renewcommand{\theequation}{B.\arabic{equation}}

In addition to the Kerr signal, we record the total (P0) and the coherence (P2) parts of the reflected optical power, as their ratio serves as a measure of the linear and/or circular birefringence. The above calculations of the ZALSI assume perfect retardance of the quarter-wave plate and absence of either linear or circular birefringence and dichroism of the sample. In reality, commercial zero-order quarter-wave plates have a typical retardance error of 1$\%$ even for normal incidence, and samples such as \CVS display birefringence and/or dichroism. As a result, the reflected beams, after passing the quarter-wave plate again, become elliptical instead of being perfectly linearly polarized. And a small fraction of the light will be incoherent with the major beams and thus won't participate in the interference. These incoherent components will not be captured by $P_1$ or $P_2$, but will still be detected as part of the average power $P_0$. And pre-factors need to be added to the formulas for $P_0$, $P_1$ and $P_2$:

\begin{align}
\centering
	P_0/P &= (1+A_0) [1+J_0(2\phi_m)]/2, \\
	P_1/P &= (1+A_1) (\sin (\varphi_{nr}) J_1(2\phi_m)), \\
	P_2/P &= (1+A_2) (\cos (\varphi_{nr}) J_2(2\phi_m))
\end{align}

where $A_0$, $A_1$ and $A_2$ are small correction pre-factors for sample birefringence and/or dichroism, and retardance error of the wave plate. The Kerr signal $\theta_K$ can be obtained using updated equation \ref {eq:Sagnac_kerr1}, with a small correction to the scaling factor. There is no change to the zero point of $\theta_K$, which is guaranteed by the symmetry of the interferometer. 

\begin{equation}
\theta_K = \frac{1}{2} \tan^{-1}\left[ \frac{(1+A_2)J_2(2\phi_m)P_1}{(1+A_1)J_1(2\phi_m) P_2}\right]
\label{eq:Sagnac_kerr1}
\end{equation}

On the other hand, a change in sample birefringence and/or dichroism will induce changes to $P_0$ and $P_2$. However, as previously mentioned, they are also dependent on $P$, which changes with focus of the objective lens and sample reflectivity. 

\begin{equation}
P_2/P_0 = \frac{(1+A_2)J_2(2\phi_m)}{(1+A_0)(1+J_0(2\phi_m))}
\label{eq:ratio}
\end{equation}

Their ratio $P_2/P_0$ is independent of these factors and represents the ratio between the coherent and the total optical powers, dubbed "coherent reflection ratio". Since the wave plate retardance error is a slow varying quantity usually dominated the slow drifts of its tilt and rotation, $P_2/P_0$ can be used to measure that change of sample birefringence and/or dichroism during temperature sweeps.

\begin{figure}[h]
	\includegraphics[width=\columnwidth]{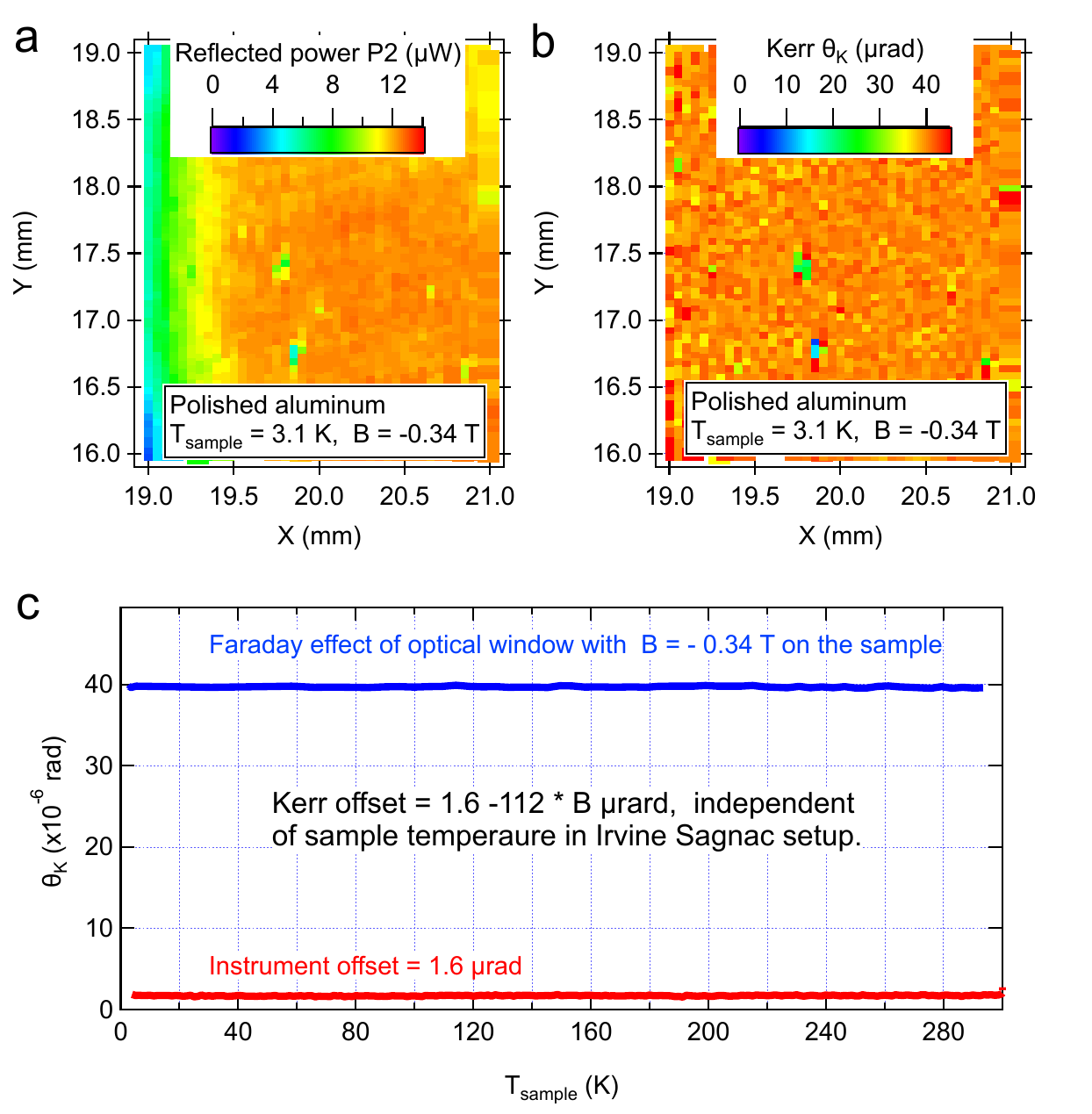}
	\caption{\textbf{Calibration of background offsets.} Window calibration of Irvine Sagnac setup. (A) P2 and (B) Kerr images of a polished aluminum surface measured at T = 3.1 $K$ and B =-0.34 T, showing a uniform background Kerr signal of 39.7 $\ \mu$rad from the optical window, even when P2 varies between 7 to 12 $\ \mu$W across the imaged region. (C) This 39.7 $\ \mu$rad optical window background (blue) as well as an instrument offset of 1.6 $\ \mu$rad are both independent of sample temperature. The total calibrated offset is 1.6 - 112 B $\ \mu$rad. }
\label{fig:background}
\end{figure}

\section{Experimental Details}\label{app:details}
\setcounter{equation}{0}\renewcommand{\theequation}{A.\arabic{equation}}

\begin{figure*}[!ht]
	\centering
	\begin{subfigure}{.48\linewidth}
		\vspace*{-2mm}
		\label{fig:Resistance_Stanford}\includegraphics[width=\textwidth]{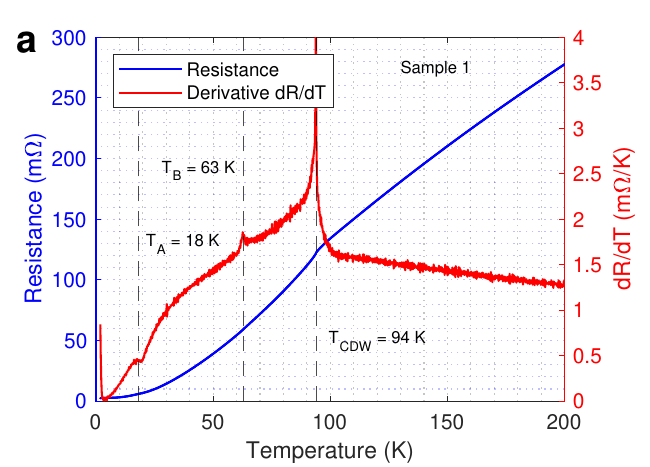}
	\end{subfigure}
	\quad
	\begin{subfigure}{.48\textwidth}
		\label{fig:Resistance_Irvine}\includegraphics[width=\textwidth]{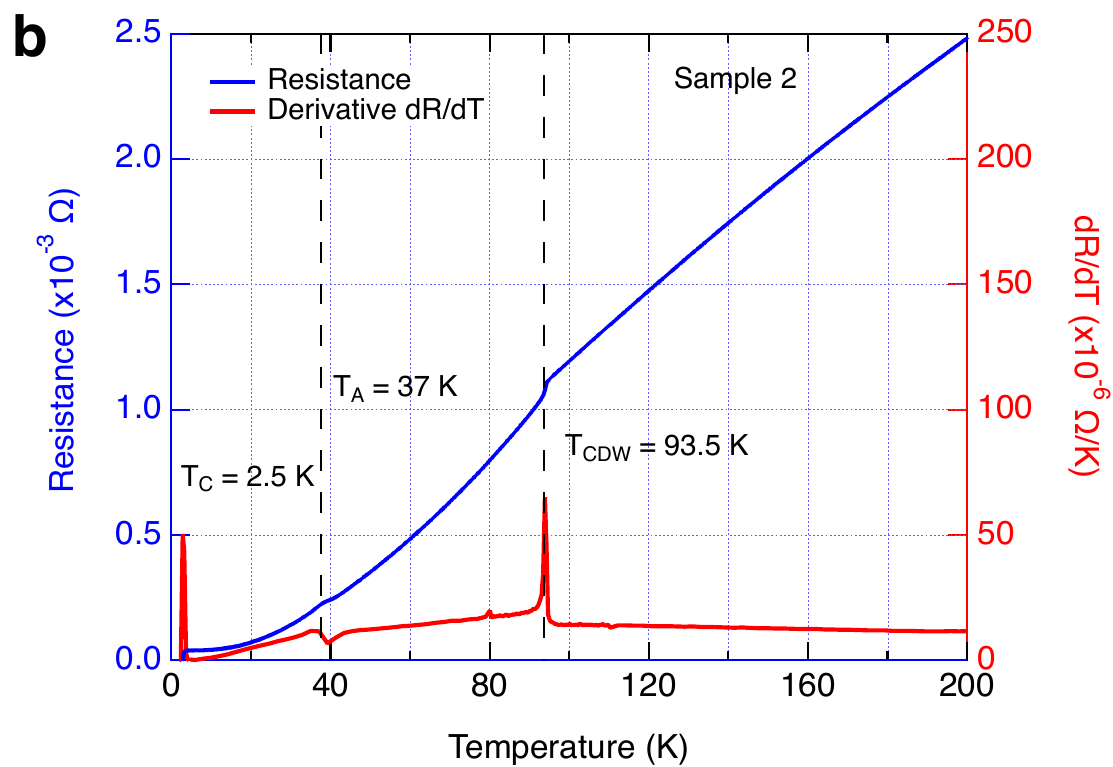}
	\end{subfigure}
	\caption{\textbf{DC resistance in CsV$_3$Sb$_5$}. 
	\textbf{a}, Resistance and derivative $dR/dT$ in CsV$_3$Sb$_5$ sample 1. \textbf{b}, Resistance and derivative $dR/dT$ in CsV$_3$Sb$_5$ sample 2. The charge density transition at $T_{CDW}$ is marked by the kink in the resistance and the pronounced peak in $dR/dT$. The superconducting transition is marked by the zero resistance below the critical temperature $T_C \approx 2.5$ K. There is an additional feature in $dR/dT$ of a dispersive line shape centered at $T_A\approx 37$ K, which is more pronouced in sample 2.}
	\label{fig:resistance_data}
\end{figure*}

In a typical experiment, we align the focused beam to an optically flat area of the crystal after cleaving. This can be facilitated by scanning imaging as shown in Fig.~\ref{fig:scan} for sample 2, where a flat region of 300 $\ \mu$m in size can be identified with high reflectivity (P0 of 10 to 30 $\ \mu$W) at T = 3.1 $K$. Kerr signals are obtained in the same scan. At B = - 0.34 T, a uniform Kerr signal of - 8.16 $\ \mu$rad is observed, even though the reflected light varies by 300$\%$ across the region. And upon removing the magnetic field, the Kerr signal is reduced to zero without any sign of spontaneous TRSB. The optical beam is then fixed at a location temperature sweeps, as presented in Fig.~2 of the main text. During zero magnetic field warmups, an absolute Kerr resolution of $30$ nrad is typically achieved over a wide temperature range between 0 and 300 $K$. For a smaller temperature range of a few $K$, we can detect a change of Kerr signal as small as $10$ nrad.

To reveal the diamagnetic shift below $T_\text{CDW} = 93.5 K$, it is necessary to perform Sagnac measurements when the sample is subject to a magnetic field. In our normal Sagnac setups\cite{PXia2006}, the fiber-optic head containing the wave-plate and focusing lens are placed inside the cryostat to achieve the lowest sample temperatures. Under a magnetic field, the fiber-optic head will contribute its own temperature-dependent Kerr signal, about $50$ $\mu\text{rad}/$T at room temperature and $100$ $\mu\text{rad}/$T below 4 $K$. Due to the long thermalization time of glass at cryogenic temperatures, the temperature of the fiber-optic head often deviates significantly from the sample temperature readings, making accurate subtractions of this background signal practically impossible during temperature sweeps. And we should be cautious with such subtractions \cite{Hu2022}. To eliminate this source of error, we have moved the optics outside the low temperature environment. In this project, both Stanford and Irvine Sagnac fiber-optic heads are located outside optical cryostats with a room temperature optical window between the focusing lens and the sample. The Kerr offset contributed by the optical window due fringing magnetic fields is thus independent of sample temperature, and can be calibrated accurately. One such calibration for the Irvine instrument is illustrated in Fig.~\ref{fig:background}, using a polished aluminum surface with negligible Kerr effect. The P2 image at T = 3.1 $K$ and B = - 0.34 T shows optical power variations between 7 and 12 $\ \mu$W across the surface, while the Kerr image taken at the same time reveals a spatially uniform Kerr background of 39.7 $\ \mu$rad due to the Faraday effect of the optical window held at 295$\pm$0.1 $K$. And we have verified that it is indeed independent of sample temperature, as shown in Fig.~\ref{fig:background}(C) (blue). There is a temperature and time-independent instrument offset of 1.6 $\ \mu$rad due to minor misalignment of optics and electronic offsets (Fig.~\ref{fig:background}(C) (red)). The total offset is thus calibrated to be 1.6 - 112 B $\ \mu$rad, which can be subtracted to produce accurate Kerr values for temperature sweeps (Fig.~2 of the main text) and spatial scans (Fig.~\ref{fig:scan}) even in a magnetic field.

\section{General Considerations for MO effects}\label{app:MO}
In general Magneto-optical (MO) effects appear because in the presence of magnetism right and left circularly polarized lights propagate differently in solids. When a magnetic field is applied to a diamagnetic insulating solid, magneto-optical effects will originate from the direct effect of the magnetic field on the orbital electronic motion. On the other hand, for ferromagnetic materials, or paramagnetic materials at low temperatures (when their Curie susceptibility is large enough), the effect of the magnetic field on the orbital motion is negligible compared with effects associated with spin-orbit interaction  \cite{Pershan1967}. For simple metals, far from plasma frequency resonances we expect that the main contribution to Kerr response is dominated by off-diagonal intraband Drude-type transitions  (i.e. originating from optical conductivity terms $\sigma_{xy}(\omega)=\sigma_0(\omega_c\tau)/[(1-i\omega\tau)^2+(\omega_c\tau)^2]$, where $\omega_c=eH/m^*c$ is the cyclotron frequency,   $ \sigma_0$ is the DC Drude conductivity and $\tau$ is the scattering time). For example, in  Al and Ag \cite{Stern1964} and nobel metals including Cu and Au \cite{McGroddy1965} these effects were measured and recently calculated, showing that for energies below $\sim 1.5$ eV the Kerr rotation is of order $\sim 10^{-9}$ rad/Oe \cite{Uba2017}. In the absence of magnetic polarization, the orbital and spin Zeeman terms will contribute off-diagonal terms through interband transitions, which for the above simple metals are at least an order of magnitude smaller. Taking into account optical and transport measurements on \CVS, both effects are expected to yield an even smaller response, which will not be detectable for the magnetic fields we used with the ZALSI experiments.

\section{DC resistance Measurements}\label{app:Resistance}
DC resistances measured on sample 1 and sample 2 are shown in Fig.~\ref{fig:resistance_data}. A kink at $T_{CDW}$ is clearly visible in the resistance of both samples, marking the charge density transition. The exact values of $T_CDW$ are determined by the temperatures of the peaks in the first derivative $dR/dT$ curve, and agree with literature data \cite{Ortiz2020} within $1$ K. The superconducting transition is marked by the zero resistance below the critical temperature $T_C \approx 2.5$ K. In the $dR/dT$ curve, especially in sample 2, there is an additional feature of a dispersive line shape centered at $T_A\approx 37$ K. We note that NMR experiments \cite{Song2022} have observed a kink behavior at a similar temperature of 35 K in $1/T_1T$, and that the electronic magneto-chiral anisotropy (eMChA) was detected in nonlinear transport \cite{Guo2022} below $34$ K. Since we observe no spontaneous Kerr signal across $T_A$, it is suggestive of a phase transition without TRSB at $T_A\approx 37$ K.

\end{document}